\newcommand\lsim{\lesssim}
\newcommand\gsim{\gtrsim}
\documentclass{emulateapj}

\begin{document}

\title{Heating and cooling of the early intergalactic medium by resonance photons}
\author{Leonid Chuzhoy \altaffilmark{1} and Paul R. Shapiro \altaffilmark{1}}

\altaffiltext{1}{McDonald Observatory and Department of Astronomy, The University of Texas at Austin, RLM 16.206, Austin, TX 78712, USA; chuzhoy@astro.as.utexas.edu; shapiro@astro.as.utexas.edu}

\begin{abstract}
During the epoch of reionization a large number of photons were produced with frequencies below the hydrogen Lyman limit. After redshifting into the closest resonance, these photons underwent multiple scatterings with atoms.
We examine the effect of these scatterings on the temperature of the neutral intergalactic medium (IGM). Continuum photons, emitted between the Ly$\alpha$ and Ly$\gamma$ frequencies, heat the gas after being redshifted into the H Ly$\alpha$ or D Ly$\beta$ resonance. By contrast, photons emitted between the Ly$\gamma$ and Ly-limit frequencies, produce effective cooling of the gas. Prior to reionization, the equilibrium temperature of $\sim 100$ K for hydrogen and helium atoms is set by these two competing processes. At the same time, Ly$\beta$ resonance photons thermally decouple deuterium from other species, raising its temperature as high as $10^4$ K. Our results have important consequences for the cosmic 21-cm background and the entropy floor of the early IGM which can affect star formation and reionization.
\end{abstract}

\keywords{intergalactic medium -- diffuse radiation -- radiation mechanisms: general}

\section{\label{Int}Introduction}
During the epoch of reionization, UV photons between the Ly$\alpha$ and Ly-limit frequencies are produced in abundance. Unlike the ionizing photons above the Ly-limit, whose mean free path in the neutral IGM is quite short, photons between Ly$\alpha$ and the Ly-limit scatter only after redshifting into one of the atomic resonances, which allows them to move freely through cosmological distances. Eventually scattering these photons exchange energy with atoms, thereby providing a mechanism for raising or lowering the gas temperature far from any radiation source.

Several authors previously considered this process. 
Madau, Meiksin \& Rees (1997) estimated that photons lose about one per cent of their energy to the gas as they redshift through 
Ly$\alpha$ resonance. They found accordingly that, if  Ly$\alpha$ scattering controls the HI hyperfine level population, as required to produce a 21-cm background distinct from the Cosmic Microwave Background (CMB), recoil heating sufficed to raise $T_{IGM}$ above  $T_{CMB}$ and to move the 21-cm background into emission.
Later, however, Chen \& Miralda-Escude (2004) showed that the inclusion of atomic thermal motion  (neglected by 
Madau et al.) in the analysis reduces the heating rate by at least three orders of magnitude. Furthermore, they showed, while 
photons emitted between Ly$\alpha$ and  Ly$\beta$ frequencies (called by them the ``continuum photons'') heat the gas when redshifted into Ly$\alpha$ resonance, those injected directly into Ly$\alpha$ by the cascade from higher resonances (called the ``injected photons''), cool the gas. 
With increasing gas temperature, the efficiency of ``continuum photons'' as heaters declines, while that of ``injected 
photons'' as coolers rises. Therefore, if, as these authors claimed, the numbers of these two types of photons were similar, then 
cooling would prevail at temperatures above 10 K. If so, the 21-cm background produced by Ly$\alpha$ pumping would have been in absorption, unless some other mechanism such as photoelectric heating by a strong X-ray background, can raise the IGM temperature above $T_{CMB}$ without fully ionizing it.

However, two important physical mechanisms were missing in previous calculations. The first is the forbidden transition from the 2s to 
1s level in hydrogen. 
The cascade which follows absorption of most photons which redshift into high atomic resonances 
(Ly$\beta$, Ly$\gamma$, Ly$\delta$ etc.) proceeds preferentially not via the 2p level (whose decay to 1s produces a Ly$\alpha$ ``injected 
photon'') but via the 2s level, instead. This significantly reduces the number of injected photons, raising the equilibrium temperature to about 100 K. The second is the impact of deuterium. Despite its very low abundance, deuterium makes an important contribution to the gas heating by its interaction with Ly$\beta$ photons. At high radiation intensities, the D Ly$\beta$ resonance becomes more important for heating the gas than H Ly$\alpha$.  In addition, scatterings by D atoms reduce the number of injected photons for hydrogen, thus decreasing the cooling rate.

In \S 2 and \S 3 we calculate the effect of resonant scattering on H and D atoms, respectively. In \S 4 and \S 5  we  discuss the impact of this scattering on the thermal evolution, 21-cm signal and reionization history of the IGM.

\section{Hydrogen resonances}
When photons redshift through the Ly$\alpha$ resonance, multiple scatterings affect their spectrum. 
The intensity $J(\nu)$ varies with frequency in the neighborhood of the resonance at $\nu_\alpha$ according to the following \cite{CS}
\begin{eqnarray}
\label{J1}
J(x)=J(0)e^{-\frac{2 \pi  \gamma  x^3}{3 a}-2 \eta  x}, %\hspace{40mm} 
\end{eqnarray}
for injected photons  and  $x>0$, otherwise
\begin{eqnarray}
J(x)= 2\pi J_0  \gamma a^{-1} e^{-\frac{2 \pi  \gamma  x^3}{3 a}-2 \eta  x} \int_{-\infty }^x e^{\frac{2\pi\gamma
   z^3}{3 a}+2 \eta  z}  z^2 \,
   dz, \hspace{0.3cm}  
\end{eqnarray}
where $a=A_{21}(2kT_{\rm H}/mc^2)^{-1/2}/4\pi\nu_\alpha$, $T_{\rm H}$ is the hydrogen kinetic temperature, $A_{21}$ is the Einstein spontaneous emission coefficient of the Ly$\alpha$ transition, $x=(\nu/\nu_\alpha-1)/(2kT_{\rm H}/mc^2)^{1/2}$ is the frequency distance from line center divided by the thermal width of the resonance, $\gamma^{-1}(1+0.4/T_{\rm s})^{-1}=\tau_{\rm GP}\gg 1$ is the Gunn-Peterson optical depth, $J_0$ is the UV intensity far away from the resonance. The recoil parameter, $\eta$, equals $[h\nu_\alpha/(2kT_{\rm H}mc^2)^{1/2}][(1+0.4/T_{\rm s})/(1+0.4/T_{\rm H})]$, where $T_{\rm s}$ is the spin temperature of the 21 cm hyperfine transition, which equals $T_{\rm H}$ ($T_{CMB}$) at high (low) radiation intensities. The intensity at $x=0$, $J(0)$, is given by
\begin{eqnarray}
\label{J3}
\frac{J(0)}{J_0}=\frac{\pi  \zeta  \left(J_{\frac{1}{3}}(\zeta
   )-J_{-\frac{1}{3}}(\zeta )\right)}{\sqrt{3}}+\,
   _1F_2\left(1;\frac{1}{3},\frac{2}{3};-\frac{\zeta
   ^2}{4}\right),
\end{eqnarray}
where $\zeta=(16\eta^3a/9\pi\gamma)^{1/2}$, $_1F_2$ is a hypergeometric function and $J_{\frac{1}{3}}$ and  $J_{-\frac{1}{3}}$ are the Bessel functions of the first kind.

\begin{figure}
\resizebox{\columnwidth}{!}
{\includegraphics{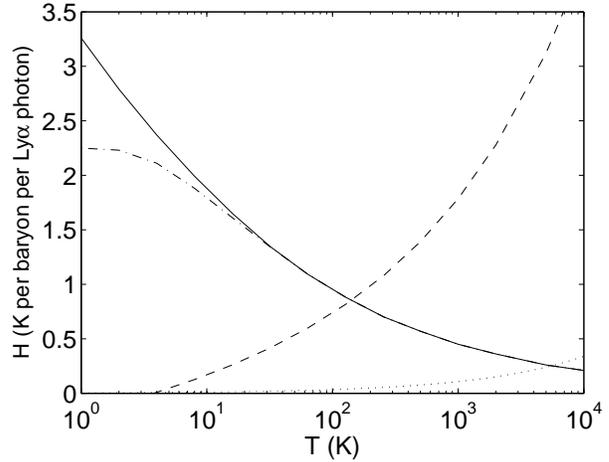}}
\caption{\label{LH}The heating/cooling rate due to continuum (solid and dash-dotted lines for $T_s=T_H$ and $T_s=T_{CMB}$ respectively) and injected (dashed line) H Ly$\alpha$ photons, and D Ly$\beta$ photons (dotted line) for flat spectrum source ($J_{\rm i}/J_{\rm c}\approx 0.15, J_{\beta}/J_{\rm c}\approx 0.35$) and $z=12$. For Ly$\alpha$ and Ly$\beta$ photons the results are plotted vs temperature of H and D atoms, respectively.}
\end{figure}

Since photons to the red (blue) of the resonance scatter preferentially off atoms moving towards (away) from them, the average energy an atom gains from a scattering photon depends on its frequency:
\begin{eqnarray}
\Delta E(x)=\frac{(h \nu)^2}{mc^2}\left(1-\frac{kT_{\rm H}}{h}\frac{\phi'(x)}{\phi(x)}\right),
\end{eqnarray}
where $\phi(x)$ is the normalized scattering cross-section \cite{CS}. The total energy gain of the gas from each photon as it passes through the resonance is determined by the radiation intensity, the probability of scattering and the average gain at each frequency:
\begin{eqnarray}
\label{etot}
\Delta E_{\rm tot}=\int \frac{J(x)}{J_0} \Delta E(x) \phi(x) dx,
\end{eqnarray}
Figure \ref{LH} shows the heating and the cooling rates obtained by numerical integration of equation (\ref{etot}).\footnote{Results are virtually insensitive to the choice of Lorentz vs Voigt profile for $\phi(x)$.} For $T_{\rm H}\gsim 100$ K and the range of optical depths corresponding to the mean density IGM  at $10<z<30$, 
$\Delta E_{\rm tot}$ can be well approximated by the asymptotes 
\begin{eqnarray}
\label{ec}
\Delta E_{\rm c}/k\sim 0.37 T_{\rm H}^{-1/3}(1+z) \; {\rm K}
\end{eqnarray}
for continuum photons and
\begin{eqnarray}
\label{ei}
\Delta E_{\rm i}/k\sim  -0.3T_{\rm H}^{1/3}(1+z)^{1/2} \; {\rm K}
\end{eqnarray}
for injected photons. The total heating/cooling rate is 
\begin{eqnarray}
\label{Hci}
H_\alpha=\dot{N_{\alpha}}(\Delta E_{\rm c}+\frac{J_{\rm i}}{J_{\rm c}}\Delta E_{\rm i}),
\end{eqnarray}
where  $\dot{N_{\alpha}}$ is the number of the photons that pass through the Ly$\alpha$ resonance per H atom per unit time.

The continuum photons (i.e., those that are redshifted into the Ly$\alpha$ resonance) can be produced in two ways. Most originate as 
photons emitted between Ly$\alpha$ and Ly$\beta$, which eventually redshift into the Ly$\alpha$ resonance. The rest come from  higher-frequency photons that are absorbed by D atoms and cascade into the D Ly$\alpha$ resonance, from which they then redshift into the H Ly$\alpha$ resonance.
The injected photons (i.e., those injected directly into the Ly$\alpha$ resonance) come from the photons emitted between Ly$\gamma$ 
and the Ly-limit. Most of the latter photons, when redshifted into the closest resonance and absorbed by H atoms, produce a cascade to 
the 2s level, from which an electron decays to the ground level by emitting two photons below Ly$\alpha$ \cite{H,PF}. Another fraction 
are absorbed by D atoms and either destroyed or converted to D Ly$\alpha$ photons, as previously explained. The rest, after redshifting 
into the closest H  resonance and producing a cascade to the 2p level, become the injected photons. Therefore, the ratio of injected 
and 
continuum photons, is
\begin{eqnarray}
\label{jic}
\frac{J_{\rm i}}{J_{\rm c}}=\sum_{n=3}^\infty \int_{\nu_n}^{\nu_{n+1}} J_{\rm s}(\nu)P_n (1-f_{\rm D,n})d\nu / \nonumber \\
 \left[\int_{\nu_2}^{\nu_3} J_{\rm s}(\nu)d\nu+\sum_{n=3}^\infty \int_{\nu_n}^{\nu_{n+1}} J_{\rm s}(\nu)P_n f_{\rm D,n}d\nu\right],
\end{eqnarray}
where $\nu_n$ is the frequency of np$\rightarrow$ 1s transition (i.e., $\nu_n=\nu_{\rm lim}(1-n^{-2})$, where $\nu_{\rm lim}=13.6\; h^{-1}$ eV is the Ly-limit frequency), $P_n$ is the fraction of all cascades from level $np$ that go through 2p (0, 0.26 for $n=3$ and 4, respectively,  and roughly 1/3 for $n>4$),  $J_{\rm s}(\nu)$ is the spectral profile of the radiation source, and $f_{\rm D,n}$ is the ratio of the cascades from level $np$ occurring in deuterium and hydrogen. 
%Finding the exact value of $f_{\rm D,n}$ requires solving several integral equations (like equation \ref{D} below), but
It can be shown that, since the deuterium optical depth, $\tau_{\rm D,n}$, in the $np\rightarrow 1s$ transition, was low  during reionization,  $f_{\rm D,n}\approx 1-e^{-\tau_{\rm D,n}(1-f_{\rm rec,n})}$, where $f_{\rm rec,n}$ is the probability for the electron in level $np$ to decay directly to  $1s$.
For a hot source with surface temperature $T_{\rm bb}\gg 10^5$ K, $J_{\rm s}(\nu)\propto \nu^2$  between Ly$\alpha$ and  Ly-limit, and $J_{\rm i}/J_{\rm c}\approx 0.17$ (the exact value depends on the optical depth of deuterium which is redshift dependent). Colder sources would produce a lower $J_{\rm i}/J_{\rm c}$ ratio, but unless the spectrum color temperature is below $\sim 5\cdot 10^4$ K, $J_{\rm i}/J_{\rm c}>0.1$.

Combining equations (\ref{ec}) and (\ref{ei}), we find that, for the mean IGM, the equilibrium temperature at which cooling by injected photons balances heating by continuum photons is \begin{eqnarray}
T_{\rm eq}\approx  130 \; {\rm K} \left(\frac{1+z}{10}\right)^{1/2} \left(\frac{J_{\rm i}/J_{\rm c}}{0.15}\right)^{-3/2} \hspace{1cm}
\end{eqnarray}
This greatly exceeds the mean temperature prior to reionization, if only adiabatic cooling occurs once the IGM decouples from the CMB.

\section{Deuterium resonances}
Unlike hydrogen, the most important resonance for deuterium is not Ly$\alpha$, but Ly$\beta$ \cite{CS}. As photons redshift through the D Ly$\beta$ resonance, a significant fraction of them are destroyed via absorption and cascade. Prior to reionization ($z\gsim 10$), the photon destruction probability is above 0.2.
Therefore the blue wing of the resonance is significantly higher than the red wing, so that the radiation color temperature around the resonance is negative.  
Because of their negative color temperature, the D Ly$\beta$ photons are much more efficient than the Ly$\alpha$ photons at heating the gas.

If the velocity distribution of deuterium atoms is Maxwellian, then the radiation intensity around the Ly$\beta$ resonance varies according to 
\begin{eqnarray}
\label{D}
\frac{e^{-x^2}}{\sqrt{\pi}}J(x)-\gamma_{\rm D} J'(x)=  \hspace{3cm}\nonumber \\
0.5f_{\rm rec,3}\int_{-\infty}^{\infty}{\rm Erfc}(Max[|x'|,|x|])J(x')dx',
\end{eqnarray}
%where 
%\begin{eqnarray}
%\gamma_{\rm D}^{-1}=2 \left(\frac{1+z}{10}\right)^{3/2}\left(\frac{n_{\rm D}/n_{\rm H}}{2\cdot 10^{-5}}\right)\left(\frac{\Omega_{\rm b} h}{0.03}\right)\left(\frac{\Omega_{\rm m}}{0.25}\right)^{-1/2}
%\end{eqnarray}
where $\gamma_{\rm D}^{-1}=\tau_{\rm D,3}\approx 3\cdot 10^3(1+z)^{3/2}(n_{\rm D}/n_{\rm H})$ \cite{CS}.
%is the optical depth of deuterium to Ly$\beta$ photons \cite{CS}.
Solving equations (\ref{etot}) and (\ref{D}) numerically, we find that for neutral IGM between $z=10$ and 20 the average energy each Ly$\beta$ photon transfers to the gas is
\begin{eqnarray}
\label{DE}
\Delta E_{\rm tot}/k\approx 0.012 T_{\rm D}^{1/2} \left(\frac{1+z}{15}\right)^{3/2} \; {\rm K},
\end{eqnarray}
where $T_{\rm D}$ is the deuterium kinetic temperature. 

Physical insight into the above result can be gained from the following argument. Photons are scattered by atoms in whose rest frame they are close to resonance (i.e., $\nu(1-v/c)\approx \nu_{\rm Ly\beta}$, where $v$ is the velocity of an atom). Thus photons in the red (blue) wing of the resonance are scattered preferentially by atoms moving towards (away) from them. When the spectrum around the resonance is flat, then to first approximation, the energy gain of atoms from photons in the blue wing is compensated by energy loss to photons on the red wing. However, when the spectrum is tilted, as in our case, each extra photon on the blue wing adds to the gas $\Delta E_{\rm tot}/k\sim(h \nu_{\rm Ly\beta}/k) v_{thermal}/c\approx 0.05 T_{\rm D}^{1/2}\; {\rm K}$, comparable to what we got in equation (\ref{DE}).
We note that, unlike H Ly$\alpha$ photons, D Ly$\beta$ photons become more efficient heaters when the temperature rises.

Since radiation affects H and D atoms differently, we now have to make separate estimates of their respective temperatures \footnote{In principle the same is also true for other species (He, $p$, $e$), but in practice only the temperature of deuterium is significantly different.}.
From equation (\ref{DE}) we find that the heating rate per D atom is
\begin{eqnarray}
\label{Hd}
H_{\beta}\approx 4\cdot 10^{-28}\;{\rm erg \cdot s^{-1}}  \times  \nonumber \\
\left(\frac{n_{\rm D}/n_{\rm H}}{2\cdot 10^{-5}}\right)^{-1} \left(\frac{\dot{N_\beta}}{10^{-14}}\right)  \left(\frac{1+z}{10}\right)^{3/2} T_{\rm D}^{1/2} ,
\end{eqnarray}
where $\dot{N_\beta}$ is the number of photons passing through the Ly$\beta$ resonance  per H atom per second.
Elastic collisions with H atoms and to a smaller extent, with He atoms, make D atoms lose energy at a rate
\begin{eqnarray}
\label{Ld}
L_{\rm D}\approx k(T_{\rm D}-T_{\rm H}) \left(\frac{3kT_{\rm D}}{2m_{\rm D}}+\frac{3kT_{\rm H}}{2m_{\rm H}}\right)^{1/2}n_{\rm H} \sigma_{\rm HD}=  \nonumber \\
1.3\cdot 10^{-31}\;{\rm erg \cdot s^{-1}}
(T_{\rm D}-T_{\rm H}) \left(\frac{T_{\rm D}}{2}+T_{\rm H}\right)^{1/2} \times \nonumber \\
\left(\frac{\Omega_{\rm b}h^2}{0.02}\right) \left(\frac{1+z}{10}\right)^3  \left(\frac{\sigma_{\rm HD}}{10^{-15}{\rm cm^2}}\right) , \hspace{1cm} 
\end{eqnarray}
where $\sigma_{\rm HD}$ is the momentum transfer cross-section for collisions between H and D atoms \footnote{Some element of uncertainty is present in the above equation due to the absence of experimental measurements of  $\sigma_{\rm HD}$.
The collisional cross sections of atoms belonging to different species can be roughly determined by the size of the atom, which for D and H atoms gives $\sigma_{\rm HD}=0.35\cdot 10^{-15} {\rm cm^2}$ \cite{CRR, Sl}. However, since D and H atoms energy levels are nearly degenerate it is possible that  $\sigma_{\rm HD}$ is significantly larger.}.
When the radiation intensity is high, so that $T_{\rm D}\gg T_{\rm H}$, $T_{\rm D}$ is set by the balance between heating and cooling rates, $L_{\rm D}=H_{\beta}$,
\begin{eqnarray}
\label{Td}
T_{\rm D}\approx 5\cdot 10^3 \;{\rm K} \left(\frac{n_{\rm D}/n_{\rm H}}{2\cdot 10^{-5}}\right)^{-1} \left(\frac{\dot{N_\beta}}{10^{-14}}\right) \times \nonumber \\
\left(\frac{\Omega_{\rm b}h^2}{0.02}\right)^{-1} \left(\frac{1+z}{10}\right)^{-3/2}  \left(\frac{\sigma_{\rm HD}}{10^{-15}{\rm cm^2}}\right)^{-1}.
\end{eqnarray}

\section{Thermal evolution of IGM}
\begin{figure}
\resizebox{\columnwidth}{!}
{\includegraphics{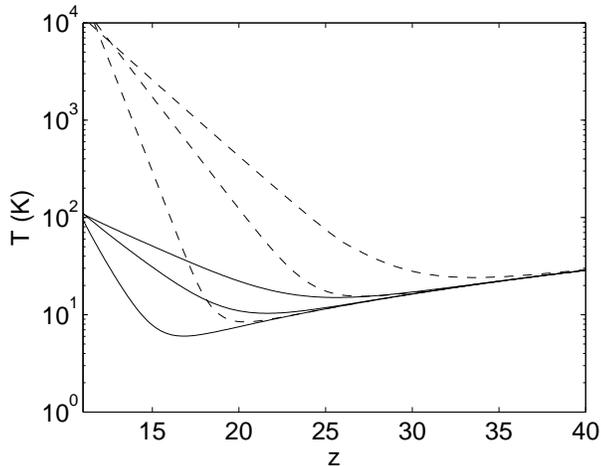}}
\caption{\label{thigm}Thermal evolution of H (solid lines) and D (dashed lines) atoms for  $\sigma_{\rm HD}=10^{-15}\; {\rm cm^2}$ and $N_{\gamma,H}(z=11)=200$. The comoving radiation intensity is assumed to grow as $e^{-z/\Delta z}$, where $\Delta z=1, 2, 3$ correspond to lower, middle and upper lines. 
%We assumed  $\sigma_{\rm HD}=10^{-15}\; {\rm cm^2}$ and $N_{\gamma,H}(z=11)=200$.
}
\end{figure}

\begin{figure}
\resizebox{\columnwidth}{!}
{\includegraphics{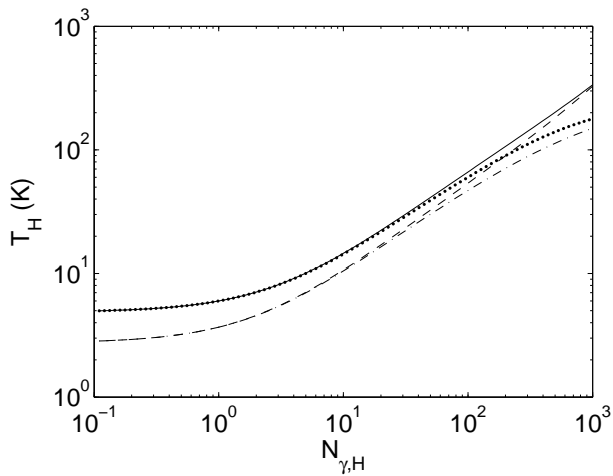}}
\caption{\label{tN}Hydrogen kinetic temperature at $z=15$ (solid and dotted lines) and $z=11$ (dashed and dashed-dotted lines) versus $N_{\gamma,H}(z)$, for $\sigma_{\rm HD}=10^{-15}\; {\rm cm^2}$. The dotted and dashed-dotted lines are the results when the heating of deuterium is neglected.}
\end{figure}

When resonance photons are the only important heating source, the temperature of H and D atoms evolves according to
\begin{eqnarray}
\frac{dT_{\rm H}}{dt}&=&\frac{2}{3k}(H_{\alpha}+L_{\rm D}\frac{n_{\rm D}}{n_{\rm H}})-\frac{4T_{\rm H}}{3t}, \\
\frac{dT_{\rm D}}{dt}&=&\frac{2}{3k}(H_{\beta}-L_{\rm D})-\frac{4T_{\rm D}}{3t}
\end{eqnarray}
where the term $4T/3t$ accounts for adiabatic cooling due to Hubble expansion. The hydrogen temperature at any redshift, $T_{\rm H}(z)$, does not depend very strongly on the history of photon production (see  Fig. \ref{thigm}). Thus $T_{\rm H}(z)$ may be expressed as a function of the total number of photons between Ly$\alpha$ and the Ly-limit per baryon emitted by redshift $z$, $N_{\rm \gamma,H}$ (Fig. \ref{tN}). If Pop III stars are the main radiation sources, then, by the end of reionization $N_{\rm \gamma,H}\sim 200$ \cite{CiM}. This implies that $T_{\rm H}$ rises to $\sim 100$ K, while $T_{\rm D}$ is raised to $\sim 10^4 (\sigma_{\rm HD}/10^{-15}\;{\rm cm^2})^{-1}$ K. However, both much higher and much lower maximum values for $N_{\rm \gamma,H}$ are not completely excluded.

\begin{figure}
\resizebox{\columnwidth}{!}
{\includegraphics{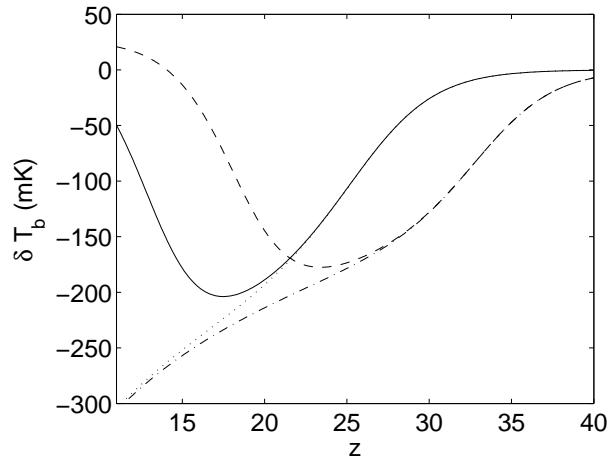}}
\caption{\label{tb1}The evolution of H 21-cm differential brightness temperature for $N_{\gamma,H}(z=11)=10$ with (solid) and without (dotted) taking the gas heating into account, and for $N_{\gamma,H}(z=11)=200$ with (dashed) and without (dashed-dotted) gas heating.
% The solid and dotted lines correspond to $N_{\gamma,H}(z=11)=10$, respectively, with and without taking the gas heating into account. The dashed and dashed-dotted lines correspond to $N_{\gamma,H}(z=11)=200$, respectively, with and without gas heating. 
The comoving radiation intensity is assumed to grow as $e^{-z/2}$ and $\sigma_{\rm HD}=10^{-15}\; {\rm cm^2}$.}
\end{figure}

\section{Implications}
We have shown that, during the reionization epoch, resonance photons can heat the neutral IGM to about $100$ K, but not much higher. 
%When gas collapses adiabatically its temperature grows with density as $\rho^{2/3}$. Since the temperature of the collapsing gas can not exceed the virial temperature, its density  can not increase by a factor greater than $(T_{vir}/T_0)^{3/2}$, where $T_0$ is the gas temperature prior to collapse. This constraint may become even tighter if the gas entropy is further increased by shockheating. 
When gas collapses out of the IGM, the entropy cannot decrease (without radiative cooling), so its overdensity inside virialized halos cannot exceed $(T_{\rm vir}/T_0)^{3/2}$, where $T_0$ $(T_{\rm vir})$ are the temperatures before (after) collapse, respectively. 
Consequently, for minihalos with $T_{\rm vir}\lsim 4000$ K that form out of gas preheated to $100$ K, the post-collapse overdensity is less than $\sim 200$.
Minihalos with such lowered central densities form $\rm{H_2}$ molecules more slowly and are more vulnerable to $\rm{H_2}$ photodissociation \cite{OH}. The photodissociation is caused by UV photons in the range from 11.6 to 13.6 eV, which overlaps the range from 10.2 to 12.7 eV (from Ly$\alpha$ to Ly$\gamma$) responsible for preheating. This may suppress $\rm{H_2}$ cooling and star formation in these minihalos.
%Since efficient ${\rm H_2}$ cooling requires even higher overdensities, preheating can help suppress star-formation in late-forming minihalos with $T_{\rm vir}\lsim 10^4$ K. By contrast, halos with $T_{\rm vir}> 10^4$ K, which cool by atomic cooling and are expected to be the primary source of ionizing photons, are unlikely to be affected. 
Minihalos would also be more easily photoevaporated, thereby reducing their consumption of ionizing photons during reionization (Shapiro, Iliev \& Raga 2004).
%In principle, therefore, by reducing gas clumpiness without significantly reducing the star-formation rate, preheating by resonance photons may actually speed up reionization.

This preheating also has a strong impact on the redshifted 21-cm signal 
from the beginning of reionization. UV resonance photons may raise $T_{\rm H}$ above $T_{\rm CMB}$, thus moving the gas from absorption to emission (see Fig. \ref{tb1}). In fact, even relatively low intensities ($\lsim 10$ photons per baryon) can drastically affect the signal. 

If the first radiation sources deposit a substantial fraction of their energy in X-rays, the average temperature of the IGM may climb even higher \cite{Ven, OH}. However, unless the X-ray spectrum is very hard, most of the energy is deposited within $\sim 1$ comoving Mpc of the source.
By contrast, the UV photons considered here, which travel cosmological distances before scattering, produce a very homogeneous heating mechanism. Therefore, in regions that do not have a radiation source nearby, heating by resonance photons can be more important even if, on average, X-ray heating is stronger.

%\section{Acknowledgment}
\acknowledgments
This work was supported by the W.J. McDonald Fellowship for LC and NASA 
Astrophysical Theory Program grants NAG5-10825 and NNG04G177G.

\end{document}